\RequirePackage{fix-cm}
\documentclass[]{fairmeta}

\usepackage{amsmath,amsfonts,bm}

\def\eqref#1{equation~\ref{#1}}

\def\1{\bm{1}}

\def\vzero{{\bm{0}}}

\def\vk{{\bm{k}}}

\def\vq{{\bm{q}}}

\def\vx{{\bm{x}}}

\def\mR{{\bm{R}}}

\DeclareMathAlphabet{\mathsfit}{\encodingdefault}{\sfdefault}{m}{sl}
\SetMathAlphabet{\mathsfit}{bold}{\encodingdefault}{\sfdefault}{bx}{n}

\usepackage{url}
\usepackage{amsmath}
\usepackage{amssymb}
\usepackage{enumitem}
\usepackage{pifont}

\definecolor{pbrule}{HTML}{4B5563}
\definecolor{pbhead}{HTML}{EEF1F5}
\definecolor{pbbody}{HTML}{FBFBFC}
\newtcolorbox{promptbox}[2][]{%
  breakable, enhanced, sharp corners, boxrule=0.5pt, colframe=pbrule,
  colback=pbbody, coltitle=black, colbacktitle=pbhead,
  fonttitle=\footnotesize\bfseries, title={#2},
  left=4pt, right=4pt, top=3pt, bottom=3pt,
  boxsep=1pt, toptitle=1.5pt, bottomtitle=1.5pt, #1}

\newcommand{\rubato}{\textsc{Rubato}}
\newcommand{\rsplit}{\textsc{Rubato}-Split}
\newcommand{\rgated}{\textsc{Rubato}-Gated}
\newcommand{\rfull}{\textsc{Rubato}-Full}
\newcommand{\singleclock}{Single-Clock}

\definecolor{qokbg}{HTML}{ECF8F3}
\definecolor{qbadbg}{HTML}{FDF1EC}
\definecolor{qbadink}{HTML}{C2410C}
\newcommand{\qbad}[1]{\textcolor{qbadink}{\textbf{#1}}}

\definecolor{qyes}{HTML}{15803D}
\definecolor{qno}{HTML}{B91C1C}
\newcommand{\cmark}{\textcolor{qyes}{\ding{51}}}
\newcommand{\xmark}{\textcolor{qno}{\ding{55}}}

\definecolor{qthinkink}{HTML}{0F766E}
\newcommand{\qthink}[1]{\textcolor{qthinkink}{\textit{#1}}}
\newcommand{\qclose}{\,{\color{qthinkink}\rule[-0.25ex]{0.7pt}{1.7ex}}\,}
\newcommand{\qnone}{{\scriptsize\color{qthinkink}(silent until the trace closed)}\ }

\newlength{\qgut}
\newcommand{\qrow}[2]{%
  \noindent\makebox[\qgut][l]{\scriptsize\itshape #1}%
  \parbox[t]{\dimexpr\linewidth-\qgut\relax}{#2}\par}
\newcommand{\qcase}[6]{%
  \noindent\colorbox{#1}{\parbox{\dimexpr\linewidth-2\fboxsep\relax}{%
    \noindent\textbf{#2}~#3\hfill{\scriptsize #4}\par\vspace{1pt}
    \qrow{Thinks}{#5}\vspace{1pt}
    \qrow{Says}{#6}}}\par\vspace{2pt}}
\newcommand{\qlegend}{%
  \noindent{\scriptsize \cmark\,judged correct\quad \xmark\,judged wrong\quad
  \qthink{spoken while still thinking}\qclose{}spoken after the trace closed\quad
  \qbad{defect} in a trace}}

\newcommand{\ua}{\textsc{ua}}
\newcommand{\sa}{\textsc{sa}}
\newcommand{\ut}{\textsc{ut}}
\newcommand{\st}{\textsc{st}}
\newcommand{\tta}{\textsc{tt}}

\title{RUBATO: Decoupled Clocks Allow Full-Duplex Models to Think as Fast as They Can}

\author[1,2,*]{Yi-Jen Shih}
\author[2]{Puyuan Peng}
\author[2]{Haohe Liu}
\author[1]{David Harwath}
\author[2]{Abdelrahman Mohamed}

\affiliation[1]{The University of Texas at Austin}
\affiliation[2]{FAIR, Meta Superintelligence Labs}

\contribution[*]{Work done at Meta}

\abstract{Recent Speech Full-Duplex (FDX) models are capable of engaging in naturalistic conversation by listening and speaking at the same time.
Building on this foundation, recent architectures introduce internal thinking streams to enhance their intelligence.
To guarantee these streams process successfully in real time, existing methods enforce a fixed decoding rate, which assumes a hardware profile capable of a specific latency. We argue this static approach is severely suboptimal. 
Because internal thinking is intrinsically asynchronous, an ideal FDX model should sustain real-time conversational flow while allowing each stream to decode at its own hardware-adaptive rate. 
Thinking streams must process as fast as the hardware allows to deliver accurate answers faster. To solve this, we propose \rubato{}, a novel multistream positional encoding that assigns independent clocks to entirely decouple the streams. This structural isolation empowers the model to zero-shot interpolate across any inference decoding speed. It drastically simplifies dataset construction by requiring training on only two extreme decoding speeds while smoothly adapting to any intermediate rate during deployment. 
\rubato{} demonstrates consistent robustness across five spoken reasoning tasks. On Spoken GSM8K, a baseline model collapses to \textbf{1.7\%} accuracy at interpolated speeds. Conversely, \rfull{} sustains \textbf{48.7\% to 51.1\%} accuracy across all rates, leveraging faster hardware to achieve a \textbf{6.5$\times$} shorter time-to-answer with a constant time-to-first-response. This advantage holds on BIG-Bench Audio and amplifies significantly during complex full-duplex interruptions. Specifically, when a user interjects while the model is actively speaking and reasoning, \rfull{} achieves \textbf{40.0\% to 45.1\%} accuracy, whereas the baseline fails completely.
}

\date{\today}
\correspondence{\email{yjshih@utexas.edu}}

\begin{document}

\maketitle

\section{Introduction}
\label{sec:intro}
Voice agents are increasingly popular as speech provides a highly intuitive medium for human-computer interaction. To unlock fluid communication, recent research focuses on Speech Full-Duplex (FDX) models~\citep{chen25_turntakingsurvey,lu26_surveyfullduplexspokendialogue}, which concurrently process incoming and outgoing audio streams in real-time to manage dynamic conversational behaviors like backchanneling and interruptions. To capture the full richness of human interaction, recent works are expanding these architectures by integrating additional streams across two primary directions. First, because natural communication relies heavily on non-verbal signals, researchers have added multimodal streams to process incoming visual cues~\citep{Jiang25_MoshiFace,lim26_faceplex,huang26_duplexomni} and generate outgoing physical actions~\citep{wang26_ellsa,zhang26_duplexsla}. Second, human conversation requires deep cognitive processing rather than just superficial spoken fluency. Because basic Speech FDX models generally lag behind their text-only counterparts in this core intelligence~\citep{xiang25_SpeechLLM_modality_gap}, researchers have also introduced internal \textit{thinking} streams. This bridges the cognitive gap, empowering the model to perform Chain-of-Thought reasoning~\citep{shih2026_canspeechLLMThink,chiang26_stitch,chiang26_shanks,wu26_chronological_think,shih26_retrothinker} and leverage external capabilities~\citep{chien2026_moshirag,arora2026_streamrag} during live conversations.

Modeling multiple streams within a Speech FDX architecture introduces a profound temporal challenge: the model must simultaneously orchestrate continuous real-time streams (e.g., audio and visual signals) alongside an internal thinking stream. 
The core complexity stems from their fundamentally incompatible pacing. 
Real-time streams operate at strict periodic intervals and require immediate generation to maintain conversational latency. In contrast, the thinking stream is triggered intermittently and is entirely decoupled from physical time, rendering it inherently asynchronous. 
To ensure the model processes all streams without violating real-time constraints, prior works~\citep{shih2026_canspeechLLMThink,zhang26_duplexsla} typically rely on a fixed synchronization strategy. By targeting a specific hardware profile, they enforce a rigid decoding pattern that allocates a fixed frame rate to the thinking stream relative to real-time inputs, guaranteeing the model won't fail during deployment.

\begin{figure}[t]
\centering
\includegraphics[width=\textwidth]{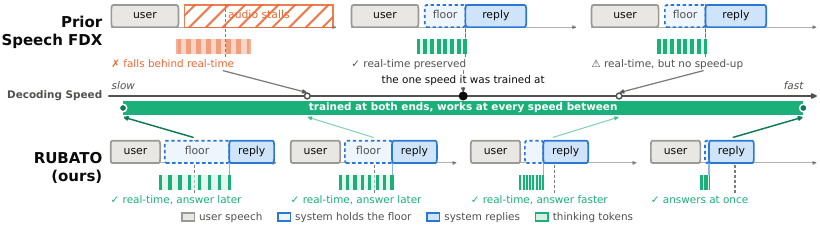}
\caption{\textbf{Comparison between \rubato{} and prior methods under varying decoding speeds.} Given an identical question and reasoning trace, varying inference speeds stretch or compress thinking tokens along the wall-clock timeline. The dotted line denotes a reference timestamp. Prior models rely on a static decoding rate where slow execution stalls real-time audio streams and fast execution wastes available compute. In contrast, \rubato{} dynamically adapts by holding the conversational floor during slower decoding and accelerating answer delivery when hardware permits.}
\label{fig:concept}
\end{figure}

We argue this static design is fundamentally limiting. An ideal Speech FDX model must allow each stream to maintain its own independent decoding pace and dynamically adapt to available compute while preserving real-time responsiveness~(Fig~\ref{fig:concept}). We motivate this decoupled approach for two key reasons. 
1) \textbf{Inelastic compute scaling:} Static decode rates for all streams prevent the model from fully leveraging fast hardware to think faster. 2) \textbf{Hardware variability:} Decoding speeds fluctuate across different GPUs, software frameworks, and dynamic server loads. This widespread variance makes it impractical to retrain the model to match every possible target decoding speed. 
Ultimately, we seek a model architecture that allows real-time interaction to remain stable while thinking streams can process as rapidly as available compute permits.

A concurrent line of research attempts to decouple heavy reasoning from real-time interaction by splitting the architecture into a fast interaction model and a slower background reasoning model~\citep{thinkingmachines26_interaction,huang26_duplexomni}. However, this dual-model paradigm introduces substantial engineering overhead for deployment and severely limits the preservation of paralinguistic audio cues during text-based communication between the two systems. More importantly, this split does not bypass the core temporal challenge. The frontend interaction model must still orchestrate real-time audio alongside the discrete tokens required to communicate with the background model. Consequently, these systems face similar bottlenecks when handling asynchronous streams while adapting to different generation speeds. Therefore, any solution that successfully eliminates these rigid temporal bottlenecks would not only enable single-model reasoning but also remain highly synergistic with dual-model architectures.

To this end, we introduce \rubato{}, a multistream positional encoding that allows multiple streams to maintain independent and adaptive decoding rates (Fig.~\ref{fig:concept}). 
We apply this to an end-to-end Speech FDX model for spoken reasoning, enabling the system to simultaneously listen, speak, and reason in real time. 
Prior works~\citep{shih2026_canspeechLLMThink,chiang26_stitch,zhang26_duplexsla} rely on a single shared position counter, a mechanism that functions properly under their fixed synchronization strategies.
Naively porting this shared counter to a system with adaptive stream pacing causes the positional distance between consecutive real-time audio tokens to be arbitrarily distorted by the dynamic volume of intermediate thinking tokens.
\rubato{} resolves this temporal distortion by assigning independent clocks to the real-time and thinking streams so that each clock advances exclusively on its designated tokens.
By doing so, \rubato{} enforces a steady temporal progression for real-time audio while allowing the thinking stream to decode as rapidly as hardware permits
\footnote{We borrow the term \rubato{} from music~\citep{hudson1994_rubato}, where the accompaniment holds a steady pulse while the melody moves freely against it, redistributing its notes without disturbing the underlying beat. This mirrors our pairing of a steady real-time audio stream with a dynamically paced thinking stream.}.

Beyond resolving this architectural bottleneck, our decoupled design significantly simplifies dataset construction and training. Simulating diverse reasoning rates in the training data severely complicates temporal alignment, as the model must gather sufficient audio context before initiating thought and conclude its reasoning before the system speaks the resulting answer. Calculating these precise alignments across varying decoding speeds imposes substantial overhead during corpus generation. \rubato{} greatly eases this data preparation process by removing the strict dependence on deployment budgets. Instead of generating data for every possible speed, we train the model exclusively on two easily generated temporal boundaries. At the baseline rate, the thinking duration aligns deterministically with the real-time audio frames. At the instantaneous rate, the entire reasoning trace resolves within a single audio frame, simulating infinite compute. Everything between these extremes acts as a zero-shot interpolation. Because \rubato{} structurally teaches the model that these timelines advance independently, the network seamlessly generalizes to any dynamic reasoning speed during deployment.

We evaluate \rubato{} on five single-turn spoken reasoning tasks from Spoken GSM8K~\citep{cobbe21_gsm8k} and BIG-Bench Audio~\citep{srivastava23_bigbench,artificialanalysis24_bigbenchaudio}, alongside complex overlapping scenarios where the model must simultaneously think, speak, and listen. 
Across all settings, \rubato{} sustains robust reasoning accuracy and a constant time-to-first-response. Crucially, it allows the system to utilize faster hardware to reduce the final time-to-answer.
In summary, our key contributions are:

\begin{itemize}
    \item \textbf{The Asynchronous-Realtime Coupling Challenge:} We identify a fundamental limitation in prior works where asynchronous thinking and real-time audio are tied together on the same timeline.
     This rigid coupling fails to accommodate dynamic generation speeds.
    \item \textbf{Novel Decoupled Clock Mechanism:} \rubato{} decouples real-time and thinking streams onto independent timelines, running efficiently on stock fused attention kernels without custom modifications.
    \item \textbf{Efficient Training via Zero-Shot Interpolation:} By structurally separating the stream timelines, \rubato{} drastically simplifies dataset construction. This design allows the model to train exclusively on two extreme temporal boundaries and generalize zero-shot to any dynamic inference speed during deployment.
    \item \textbf{Robust Empirical Performance:} We demonstrate that our framework sustains high reasoning accuracy across dynamic inference speeds on complex spoken reasoning tasks. Crucially, it directly translates faster hardware into significantly reduced user-facing latency.
\end{itemize}

\section{Background}
\label{sec:background}
\subsection{RoPE: Rotary position embeddings}
\label{sec:bg_rope}
An attention head of dimension $D$ maps the token at sequence position $m$ to a query $\vq_m \in \mathbb{R}^{D}$ and a key $\vk_m \in \mathbb{R}^{D}$.
Rotary position embeddings (RoPE)~\citep{su21_rope} inject position by rotating both vectors in $D/2$ two-dimensional planes before the dot product, plane $i$ turning at angular frequency
\begin{align}
\theta_i \;=\; \Theta^{-2(i-1)/D}, \qquad i = 1, \dots, D/2 ,
\label{eq:rope_freqs}
\end{align}
with frequency base $\Theta$.
Position $m$ acts through the block-diagonal matrix $\mR(m)$ whose $i$-th $2\times 2$ block rotates by the angle $\theta_i m$, and because $\mR(m)^{\!\top}\mR(n) = \mR(n-m)$, the attention logit between a query at $m$ and a key at $n$ is
\begin{align}
\big\langle \mR(m)\,\vq_m,\; \mR(n)\,\vk_n \big\rangle \;=\; \big\langle \vq_m,\; \mR(n-m)\,\vk_n \big\rangle ,
\label{eq:rope_relative}
\end{align}
which depends on the two positions only through their offset $n-m$.
In implementation, the rotation is applied to the query and the key independently.
Attention therefore stays a plain dot product over the rotated tensors and can be computed by a fused kernel such as FlashAttention~\citep{Dao22_flashattention}.

\subsection{MultiStream Full-Duplex Models}
\label{sec:bg_full-duplex}

To enable full-duplex interaction, we adopt the four-stream interleaving strategy proposed in SCoT~\citep{arora25_cot_e2e_speech_fullduplex}. 
This architecture consists of User Audio (\ua{}), System Audio (\sa{}), User Text (\ut{}), and System Text (\st{}). 
By providing time-aligned ASR and TTS text for the user and system, respectively, they improve semantic coherence. During preprocessing, audio tokenization compresses the waveforms into discrete tokens, where the temporal span of a single token defines one \emph{frame} (e.g., 80\,ms).

Prior architectures model these streams via concatenation or interleaving. Concatenation~\citep{kyutai24_moshi,roy26_personaplex} forces a synchronized processing rate, while interleaving~\citep{wang26_ellsa,zhang25_OmniFlatten} flattens streams into a single 1D sequence for greater flexibility. Both paradigms face strict real-time constraints, meaning the decoding computation must finish faster than the physical duration of the processed audio frames. Because modern GPUs decode rapidly, the residual compute time within each interleaved chunk can be leveraged to decode internal thinking streams like reasoning or RAG.

We adopt the interleaved paradigm due to its flexibility. Because hardware decoding speeds dictate how many reasoning tokens can be generated between consecutive real-time audio frames, we formally define this dynamic capacity as the thinking budget $\mathcal{B}$. Let $\ua_f, \sa_f, \ut_f, \st_f$, and $\tta_{f,k}$ represent the tokens for their respective streams at frame $f$. We capture frame-level hardware fluctuations by assigning a dynamic budget $\mathcal{B}_f$ to frame $f$, defining the interleaved chunk $C_f$ as:
\begin{equation}
C_f = \left[ \ua_f, \sa_f, \ut_f, \st_f, \tta_{f,1}, \dots, \tta_{f,\mathcal{B}_f} \right]
\label{eq:interleave}
\end{equation}
Real-time tokens are positioned first to ensure immediate processing, followed by the variable sequence of thinking tokens. The model consumes these chunks sequentially as a single flat sequence $\vx = [\,C_1, \dots, C_F\,]$, and we use $t$ to denote the position of a token in the flattened sequence.
Since $\mathcal{B}$ is a per-frame limit, any reasoning process exceeding the current budget simply overflows into the next frame. Unlike prior works that fix $\mathcal{B}$ during training and inference, our framework allows $\mathcal{B}$ to depend on inference time capacities. 
In practice, $\mathcal{B}$ depends on model size, codec frame rate, serving techniques, and hardware. To simplify, we evaluate varying $\mathcal{B}$ directly and defer detailed estimates to Appendix~\ref{sec:gpu_budget}.

\textbf{The Deficit of Single-Clock.} Existing methods apply RoPE to this flattened sequence using a single monotonically increasing index. While effective for a fixed $\mathcal{B}$, this design creates a positional distribution shift when inference budgets differ from those observed during training.
Dynamically expanding or shrinking the intermediate thinking tokens arbitrarily distorts the positional distance between consecutive real-time tokens like $\ua_f$ and $\ua_{f+1}$. This single-clock constraint thereby prevents the model from robustly learning the true temporal alignment of the streams.

\section{Proposed Method}
\label{sec:method}
\begin{figure}[t]
\centering
\includegraphics[width=\textwidth]{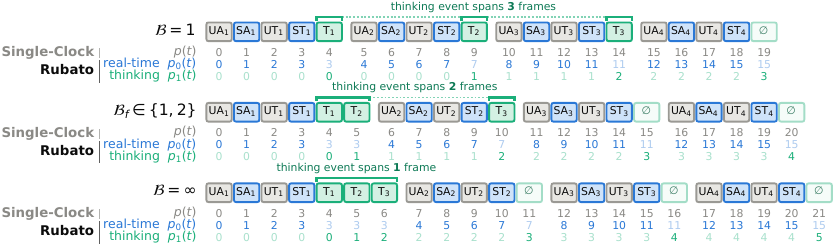}
\caption{\textbf{Positional index arrangement for \rubato{} vs. Single-Clock. } 
An example 3-token thinking event ($\mathrm{T}_1$ to $\mathrm{T}_3$) starting at frame $1$ spans a variable number of frames based on $\mathcal{B}_f$. 
While Single-Clock applies a continuous index that arbitrarily distorts the distance between real-time tokens, \rubato{} assigns independent clocks that advance exclusively on designated tokens (full color) and pause elsewhere (faint). 
This guarantees consecutive real-time tokens remain strictly equidistant, fully decoupling the audio and reasoning timelines.
}
\label{fig:token_arrangement}
\end{figure}

\subsection{\rubato{}}
\label{sec:clocks}

To overcome the limitations of a single clock, we introduce \rubato{}, which assigns multiple independent positional indices (clocks) to the FDX model's streams. While extendable to $C$ clocks, we use two for our setup: a real-time clock ($c=0$) and a thinking clock ($c=1$). We write $\kappa(t) \in \{0, 1\}$ for the clock of the token at flat position $t$, which its stream fixes, so that $\kappa(t) = 0$ on the four real-time streams \ua{}, \sa{}, \ut{} and \st{}, and $\kappa(t) = 1$ on the thinking stream \tta{}.

Clock $c$ advances exclusively when processing its assigned tokens. Let $n_c(t)$ denote the total number of clock-$c$ tokens up to sequence position $t$:
\begin{equation}
n_c(t) = \sum_{u \leq t} \mathbf{1}\big[\kappa(u) = c\big], \qquad p_c(t) = \max\big(n_c(t) - 1, 0\big).
\label{eq:clock_positions}
\end{equation}
Here, $p_c(t)$ represents the zero-based index of the most recent clock-$c$ token at or before position $t$. Because a clock pauses on unassigned tokens, every token carries a $C$-dimensional position vector $\big(p_0(t), \dots, p_{C-1}(t)\big)$ rather than a scalar index (Fig.~\ref{fig:token_arrangement}). 
This structural decoupling successfully desynchronizes the two timelines. 
It guarantees $p_0(\ua_f) = 4(f-1)$ regardless of the intermediate thinking budget, while simultaneously ensuring that the thinking tokens maintain a continuous, uninterrupted index sequence completely unaffected by the real-time frames interleaved among them.
Next, we integrate these clocks into attention via three methods, letting $m$ and $n$ index query and key positions in the flattened sequence.

\subsubsection{\rsplit{}}
\label{sec:rubato_split}
The most direct approach to applying multiple positional indices is to adapt TM-RoPE~\citep{xu25_qwen25} by giving each clock its own subspace of the rotary dimensions.
Formally, \rsplit{} partitions the $D$ rotary coordinates into $C$ contiguous subspaces of $D/C$ coordinates each, writes $\vq_m = [\,\vq_{m,0} ; \dots ; \vq_{m,C-1}\,]$ for the query's components in them and the same for $\vk_n$, and rotates subspace $c$ by clock $c$'s position,
\begin{align}
\tilde{\vq}_{m,c} \;=\; \mR^{(C)}\big(p_c(m)\big)\,\vq_{m,c} ,
\qquad\qquad
\tilde{\vk}_{n,c} \;=\; \mR^{(C)}\big(p_c(n)\big)\,\vk_{n,c} .
\label{eq:split}
\end{align}
Here $\mR^{(C)}$ is the rotation of Sec.~\ref{sec:bg_rope} carried over to one subspace, block diagonal over the $D/(2C)$ planes that subspace holds, with its $i$-th block turning at
\begin{align}
\theta^{(C)}_i \;=\; \Theta^{-2C(i-1)/D} \;=\; \theta_{C(i-1)+1} ,
\qquad i = 1, \dots, D/(2C) .
\label{eq:split_freqs}
\end{align}
Consequently, each subspace retains the same rotation frequency range while diluting the resolution by $C$. 
At $C=1$, the construction reduces back to $\mR^{(1)} = \mR$.
The attention score is the sum across each subspace dot product with dedicated rotation:
\begin{align}
\big\langle \tilde{\vq}_m, \tilde{\vk}_n \big\rangle \;=\; \sum_{c=0}^{C-1} \big\langle \vq_{m,c},\; \mR^{(C)}\big(p_c(n) - p_c(m)\big)\,\vk_{n,c} \big\rangle .
\label{eq:split_logit}
\end{align}
Every pair is therefore scored by every clock at once. 
An audio-to-audio pair still carries a thinking-clock term, and a thinking-to-thinking pair still carries a real-time one, but in different subspaces.

\subsubsection{\rgated{}}
\label{sec:rubato_gated}
A primary drawback of \rsplit{} is that it indiscriminately applies all clock rotations regardless of the query and key token types, potentially injecting noise into the attention mechanism. 
For example, the attention between two \ua{} tokens should strictly depend on their distance along the real-time clock. 
Under \rsplit{}, a varying number of intermediate thinking tokens causes irrelevant rotations within the thinking clock subspace.

To eliminate this noise, we propose \rgated{} to apply subspace rotations conditionally. 
To determine the exact rules, we analyze two query-key scenarios:
1) \textit{Same Clock}: We naturally apply their shared clock's rotation.
2) \textit{Different Clocks}: We apply the key token clock's rotation, allowing the query to differentiate this specific key from other keys sharing that same clock.
Consolidating these cases reveals that the target rotation always matches the key token. 
We can thus simplify the mechanism into a single unified rule: rotate solely based on the key token's clock.
Formally, letting $c^\star = \kappa(n)$ denote the clock assigned to the key token, the attention score is computed as:
\begin{equation}
\big\langle \tilde{\vq}_m, \tilde{\vk}_n \big\rangle = 
\big\langle \vq_{m,c^\star}, \mR^{(C)}\big(p_{c^\star}(n) - p_{c^\star}(m)\big) \vk_{n,c^\star} \big\rangle
+ \sum_{c \neq c^\star} \big\langle \vq_{m,c}, \vk_{n,c} \big\rangle.
\label{eq:gated}
\end{equation}

\subsubsection{\rfull{}}
\label{sec:rubato_full}
Building on the concept of \rgated{}, we observe that only a single clock dictates the rotation for any given query-key pair. We can therefore use that exact clock to rotate all dimensions, thereby preventing any frequency resolution dilution. We define this method as \rfull{}.
Formally, the attention computation remains identical to Eq.~\ref{eq:rope_relative}, except it applies the rotation using the key token's clock, denoted as $c^\star = \kappa(n)$:
\begin{equation}
\big\langle \tilde{\vq}_m, \tilde{\vk}_n \big\rangle = \big\langle \vq_m, \mR\big(p_{c^\star}(n) - p_{c^\star}(m)\big) \vk_n \big\rangle.
\label{eq:full_logit}
\end{equation}

\subsubsection{\rubato{} Implementation}
\label{sec:rubato_impl}

\textbf{Training.} For \rsplit{}, the implementation is simple. We apply different clocks to separate rotary subspaces before calling FlashAttention. For \rgated{} and \rfull{}, the rotation depends on the key token's clock. To utilize FlashAttention without writing custom Triton kernels, we expand the hidden dimension to fold this dependence into the tensors. The query pre-computes all potential clock rotations side-by-side. The key retains only the rotation for its specific clock and zeroes out the rest. Consequently, the dot product natively isolates the matching clock pair.

\rgated{} applies this logic per subspace, reusing the rotated components from Eq.~\ref{eq:split}:
\begin{equation}
\begin{array}{r@{\,}c@{\,}c@{\,}c@{\,}c@{\,}c@{\,}c@{\,}c@{\,}c@{\,}c@{\,}c@{\,}c@{}l}
\tilde{\vq}_m = [ & \tilde{\vq}_{m,0} & ,\dots, & \tilde{\vq}_{m,c^\star} & ,\dots, & \tilde{\vq}_{m,C-1} & , &
                    \vq_{m,0} & ,\dots, & \vq_{m,c^\star} & ,\dots, & \vq_{m,C-1} & ] \\[3pt]
\tilde{\vk}_n = [ & \vzero & ,\dots, & \tilde{\vk}_{n,c^\star} & ,\dots, & \vzero & , &
                    \vk_{n,0} & ,\dots, & \vzero & ,\dots, & \vk_{n,C-1} & ]
\end{array}
\label{eq:gated_impl}
\end{equation}

\rfull{} applies this to the entire head by duplicating it $C$ times:
\begin{equation}
\begin{array}{r@{\,}c@{\,}c@{\,}c@{\,}c@{\,}c@{\,}c}
\tilde{\vq}_m = [ & \mR\big(p_0(m)\big)\vq_m & ,\dots, & \mR\big(p_{c^\star}(m)\big)\vq_m & ,\dots, & \mR\big(p_{C-1}(m)\big)\vq_m & ] \\[3pt]
\tilde{\vk}_n = [ & \vzero & ,\dots, & \mR\big(p_{c^\star}(n)\big)\vk_n & ,\dots, & \vzero & ]
\end{array}
\label{eq:full}
\end{equation}

Because zero-valued key elements nullify unselected query rotations, the dot product perfectly recovers Eq.~\ref{eq:gated} and Eq.~\ref{eq:full_logit}. 
This approach is simple to implement during training and doesn't cause too much burden, provided the expanded dimension $C \times D$ remains relatively small.

\textbf{Inference.} Because decoding scores a single query without a backward pass, we drop the dimension expansion training trick to maximize speed. Preserving decoding efficiency is crucial because any latency overhead directly degrades the thinking budget $\mathcal{B}$. Conceptually, we maintain $C$ independent KV caches. Each key is rotated by its designated clock and stored exclusively in that cache. During generation, we create $C$ rotated copies of the incoming query, score each against its corresponding cache, and merge the $C$ partial softmax results using the log-sum-exp values returned by the kernel~\citep{ye25_flashinfer}. This guarantees exact mathematical equivalence to the training path while preserving strict parity with Single-Clock decoding speeds. Details are available in Appendix~\ref{sec:multi_kv}.

\subsection{Training on the two extreme budgets}
\label{sec:two_extremes}

The structural design of \rubato{} inherently allows us to interpolate $\mathcal{B}$, which simplifies the effort of compiling training data.
Specifically, we propose training the model on just two boundary conditions: $\mathcal{B}=1$ and $\mathcal{B}=\infty$. 
Let $\mathcal{T}=(l, s)$ represent a thinking event comprising $l$ tokens that begins at frame $s$. 
For $\mathcal{B}=1$, the event strictly requires $l$ frames to execute. 
This deterministic duration allows us to properly align the reasoning process with the real-time conversation or insert verbal audio fillers to preserve system responsiveness, similar to how people hold the floor when processing complex questions.
For $\mathcal{B}=\infty$, the model interleaves the entire $l$-token thinking trace instantly between frame $s$ and frame $s+1$. 
This effectively decouples the reasoning process from the wall-clock time and largely reduces the alignment complexities between the audio streams.

\section{Experiments}
\label{sec:evaluation}
\subsection{Training and Evaluation Data}
\label{sec:data}

We evaluate our approach on GSM8K~\citep{cobbe21_gsm8k} and BIG-Bench Audio~\citep{srivastava23_bigbench,artificialanalysis_bba} for two primary reasons. First, they are commonly used benchmarks. Second, they comprise spoken mathematical and logical tasks where intermediate reasoning is critical for accuracy. Solving these tasks successfully thus provides strong evidence that the model's real-time and thinking streams are coordinating effectively. 
Because no suitable full-duplex spoken reasoning corpora exist for training, we synthesize three new datasets via Qwen3 TTS~\citep{Qwen3-TTS}. The first two, \emph{Spoken GSM8K} and \emph{Spoken BIG-Bench Hard}, are single-turn datasets derived from the GSM8K-Aug-NL~\citep{wu25_gsm8k_augnl} and BIG-Bench Hard~\citep{suzgun23_bigbenchhard} text corpora to include mainstream math and logical reasoning. 
We rigorously make sure there is no test set leakage.

For these single-turn datasets, we generate balanced training examples mapped to the two thinking budget extremes. For $\mathcal{B}=1$, the system responds to the user's question with a spoken filler designed to hold the floor. This filler's duration exactly matches the frame count required to complete the underlying thinking trace before speaking the final answer. For $\mathcal{B}=\infty$, the system skips the filler and answers immediately since all reasoning computes within a single frame.

Our third dataset introduces a highly challenging overlapped scenario by randomly merging two \emph{Spoken GSM8K} QA pairs into one conversation. This setup demands that the model actively reason and speak its answer to the first question while simultaneously listening to the user interject with a second question. Further dataset details are provided in Appendix~\ref{sec:data_details}.

\textbf{Metrics.}
Spoken answers are transcribed and then judged by Deepseek V4 Pro~\citep{deepseekv4}, which reads the reference answer and decides whether the model committed to the correct value. 
For the overlapped benchmark, the judge scores each question separately.
In addition to accuracy, we report two latencies, both measured from the moment the user stops speaking. \emph{Time to first response} (TTFR) is the onset of any system speech, and \emph{time to answer} (TTA) is the onset of the spoken final answer.
The two come apart exactly when the model speaks a filler: TTFR is when it starts saying something, and TTA is when it commits.

\textbf{Inference Thinking Budget Sweep.} To assess interpolation robustness, we evaluate multiple fixed and random thinking budgets between the $1$ and $\infty$ training extremes (marked by bold ticks in all figures). 
We test six fixed budgets $\mathcal{B} \in \{1, 2, 4, 8, 16, 32\}$, the $\mathcal{B}=\infty$ upper bound, and three \emph{random} configurations. 
In these random settings, the per-frame budget $\mathcal{B}_f$ is sampled uniformly from the intervals $\left[1, 2\right]$, $\left[2, 3\right]$, or $\left[3, 4\right]$ and plotted at the expected mean. Results for highly erratic random budgets like $\left[1, 16\right]$ are deferred to Appendix~\ref{sec:extra_budgets} as they rarely occur in real-world deployments.

\subsection{Model Implementation}
\label{sec:implementation}

All models start from the Gemma-4 E4B-IT~\citep{gemmateam26_gemma4} text backbone and are trained as one autoregressive model over the flat sequence of Eq.~\ref{eq:interleave}.
Audio is tokenised with NanoCodec at 12.5\,Hz~\citep{casanova25_nanocodec}, giving 80\,ms frames, into 4 codebooks of 4032 entries.
Frames with no thinking content predict the \texttt{<empty-think>} token.
On inference, after finishing decoding a frame's real-time tokens, we start decoding the thinking tokens until either $\mathcal{B}_f$ is reached or the model samples \texttt{<empty-think>} token.
Other training and inference details are in Appendix~\ref{sec:hyper_params}.

\subsection{Single-Turn Spoken Reasoning}
\label{sec:single_turn}

\begin{figure}[t]
\centering
\includegraphics[width=\textwidth]{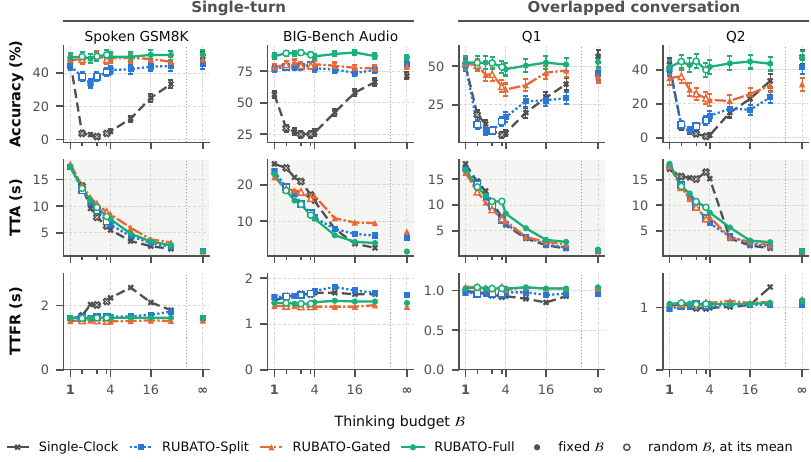}
\caption{\textbf{Spoken reasoning performance across varying inference thinking budgets.} Models are trained exclusively on the two extreme rates ($\mathcal{B}=1$ and $\mathcal{B}=\infty$, marked by bold ticks), meaning all intermediate points evaluate zero-shot interpolation. Filled and hollow markers denote fixed and random per-frame budgets (plotted at their mean) respectively. Panels display 95\% Wilson intervals for accuracy and medians for latency. We report Accuracy (Acc.), Time-to-Answer (TTA), and Time-to-First-Response (TTFR) on two single-turn benchmarks, alongside isolated results for both questions in the overlapped setting.}
\label{fig:single_overlap}
\end{figure}

We jointly train our models on the Spoken GSM8K and Spoken BIG-Bench Hard and evaluate them on Spoken GSM8K test set and BIG-Bench Audio independently. The first two columns of Fig.~\ref{fig:single_overlap} highlight several critical findings:

\textbf{\rubato{} excels at interpolating unseen thinking budgets.} All three \rubato{} variants outperform the \singleclock{} baseline consistently on both benchmarks. They maintain high accuracy and a stable time-to-first-response alongside a strictly decreasing time-to-answer as the budget scales. Conversely, \singleclock{} accuracy collapses to under 10\% on Spoken GSM8K and roughly 25\% on BIG-Bench Audio at several decoding speeds. This confirms that \rubato{} empowers the model to dynamically exploit available inference compute without sacrificing responsiveness. It also validates that \rubato{} preserves both intra- and inter-stream correctness, prompting the answer immediately after reasoning finishes.

\textbf{\rfull{} and \rgated{} deliver the highest stability among \rubato{} variants.} Measuring maximum accuracy drops from training endpoints to interpolated budgets, \singleclock{} and \rsplit{} degrade by 45.0 and 10.6 points on Spoken GSM8K, whereas \rgated{} and \rfull{} drop only 1.1 and 0.6 points. BIG-Bench Audio follows a similar trend. By only rotating the key token's clock for every query-key pair, \rgated{} and \rfull{} completely eliminate the cross-clock contamination seen in \rsplit{}. Additionally, \rfull{} outperforms \rgated{} by 6.3 to 12.2 points on BIG-Bench Audio. This advantage is likely because preserving the vanilla RoPE frequency basis avoids frequency dilution and aligns perfectly with the pretrained Gemma backbone.

\subsection{Overlapping-Turn Spoken Reasoning}
\label{sec:overlap_turns}

We additionally train and evaluate our model on Overlapped Spoken GSM8K, a demanding scenario requiring simultaneous speaking, listening, and thinking. In this challenging setup, the user asks an initial question and subsequently interjects with a second question right as the system is actively thinking and speaking.
Based on the last two columns of Fig.~\ref{fig:single_overlap}, we draw the following conclusions:

\textbf{\rfull{} widens its performance gap under complex full-duplex conditions.} 
\singleclock{} and \rsplit{} collapse entirely on both questions. They only get 1 to 14\% correct around $\mathcal{B}$ at 2 to 4. In contrast, \rfull{} stays within a few points of its trained endpoints and correctly answers the interrupting question in 40.0\% to 45.1\% of cases across every unseen rate. The most telling comparison is \rgated{}. 
While \rgated{} underperforms \rfull{} by a mere 1.6 points on average during single-turn GSM8K evaluations, this gap widens roughly five-fold on Q1 and ten-fold on Q2 in the overlapped setting. 
Overlapping speech thus reveals exactly where clock construction choices become more critical. 
Furthermore, \rfull{} retains its latency benefits under these overlapping conditions, reducing the median time-to-answer by a factor of six from $\mathcal{B}=1$ to $\mathcal{B}=32$ on both questions.

\textbf{\rfull{} minimizes the interruption effect on the second question.} For every method and budget, Q1 accuracy exceeds Q2, reflecting the inherent hardness of absorbing a new question while actively answering a previous one. 
Among the architectures that survive interpolation, \rfull{} suffers an average drop of only 7.8 points from Q1 to Q2, compared to a 15.6-point drop for \rgated{}. 

\subsection{Spoken Filler Quality}
\label{sec:filler_quality}

\begin{figure}[t]
\centering
\includegraphics[width=\textwidth]{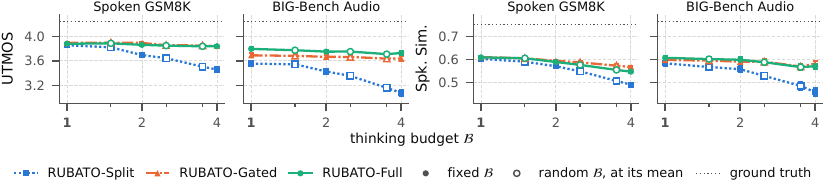}
\caption{\textbf{Spoken-filler quality across thinking budgets.}
To evaluate the quality of the system audio filler, we report the UTMOS score and speaker similarity. From the figure we see that \rfull{} and \rgated{} maintain the audio quality while \rsplit{} degrades significantly as the thinking budget is interpolated.}
\label{fig:filler_scores}
\end{figure}

Although accuracy and latency metrics demonstrate the core effectiveness of \rubato{}, we must also verify that interpolating the inference thinking budget preserves generated audio quality. To assess this, we measure UTMOS and Speaker Similarity scores specifically within the spoken filler regions, where interleaved thinking tokens are most likely to introduce acoustic artifacts. We omit the \singleclock{} variant due to its prior catastrophic failure during interpolation, and we limit our evaluation to budgets of $1 \leq \mathcal{B} \leq 4$ because higher budgets generate fillers that are too short for reliable scoring.

As shown in Figure~\ref{fig:filler_scores}, \rsplit{} suffers significant degradation in both UTMOS and Speaker Similarity during interpolation despite its relatively strong accuracy in Figure~\ref{fig:single_overlap}. 
Conversely, \rgated{} and \rfull{} remain notably stable. This reinforces our conclusion that eliminating noise from unrelated clocks substantially improves architectural robustness when processing multistream sequences at varying rates.

\subsection{Training-Layout Mixture}
\label{sec:ablations}

We additionally perform an ablation on the ratio of $\mathcal{B}=1$ examples within the training data (detailed in Appendix~\ref{sec:ablation_filler}). \rfull{} proves significantly more robust than \rgated{} across varying dataset mixtures. It sustains accuracy at the trained rate between 41.6\% and 53.7\% over the five corpora, while \rgated{} experiences a performance spread nearly three times wider (20.5\% to 52.7\%). 
Despite this disparity, a balanced distribution between the $1$ and $\infty$ budgets delivers optimal performance for both architectures. Qualitative model outputs are provided in Appendix~\ref{sec:qualitative_examples}.

\section{Conclusion}
\label{sec:conclusion}
In this paper, we address the fundamental challenge of training end-to-end Speech FDX models that can freely reason and maintain responsiveness while dynamically adapting to variable inference hardware speeds. We propose \rubato{}, a novel positional encoding framework that explicitly models the temporal relationships within and across multiple data streams. This architectural decoupling fundamentally simplifies the training development pipeline. A model only needs to be trained once using two extreme boundaries, yet it effortlessly deploys across diverse generation speed by enabling zero-shot interpolation to any dynamic real-time decoding rate. Our empirical results on complex spoken reasoning benchmarks demonstrate that our \rfull{} architecture effectively leverages available compute to accelerate responses without sacrificing accuracy. Ultimately, the impact of \rubato{} extends far beyond Speech FDX. The core mechanism of independent positional clocks provides a potential solution for any architectural scenario requiring the orchestration of multiple asynchronous token streams or dynamic generation rates, paving the way for more resilient and flexible multistream models.

\subsection*{AI use statement}

Every contribution in this paper was motivated, designed, and carried out by the authors, who take full responsibility for all of its results and conclusions.
We did not use generative AI tools for research ideation or for literature search.
We used Claude Code as a coding assistant, both for implementing and refactoring parts of the training, inference, and evaluation code and for writing the scripts that render the figures and tables, and we used Gemini and Codex to polish the prose.
We treated everything these tools produced as a draft to be verified rather than as a result to be accepted, and we checked all of it.
AI-assisted code was read and tested by the authors, and its numerically load-bearing parts carry their own checks, since the decode-time attention path is verified against the reference implementation in Appendix~\ref{sec:multi_kv} and every figure is regenerated from the recorded evaluation runs.
Every claim and every number in the text was traced back to those runs by an author, and any statement we could not source that way was removed rather than kept.
Generative models also appear inside the method itself, synthesising the corpora and scoring the outputs, and those uses are named and described in Sec.~\ref{sec:data} and Appendix~\ref{sec:data_details}.

\subsection*{Reproducibility statement}

The three attention constructions are given in closed form in Sec.~\ref{sec:method}, as Eq.~\ref{eq:split} for \rsplit{}, Eq.~\ref{eq:gated} for \rgated{} and Eq.~\ref{eq:full_logit} for \rfull{}, together with the training-time expansions in Eqs.~\ref{eq:gated_impl} and \ref{eq:full} that let all three run on a stock fused attention kernel; the decode-time construction and its equivalence check are in Appendix~\ref{sec:multi_kv}.
The clock assignment these operate on is defined in Sec.~\ref{sec:clocks} and the flat sequence layout in Eq.~\ref{eq:interleave}.
Backbone, tokeniser, objective, optimiser, and decoding configuration are in Sec.~\ref{sec:implementation} and Appendix~\ref{sec:hyper_params}, and the two-budget training recipe that the corpora are built around is in Sec.~\ref{sec:two_extremes}.
Corpus construction, sizes, and splits are in Sec.~\ref{sec:data} and Appendix~\ref{sec:data_details}, and that same section states the judge configuration and the two latency definitions.
Appendix~\ref{sec:extra_budgets} reports the one budget setting the figures leave out.

\clearpage
\newpage
\bibliographystyle{assets/plainnat}
\bibliography{references}

\clearpage
\newpage
\beginappendix
\section{Corpus construction and statistics}
\label{sec:data_details}
\textbf{Overview of Training Datasets.}
Table~\ref{tab:corpus_budgets} details the statistics of our two primary constructed training datasets: \emph{Spoken GSM8K} and \emph{Spoken BIG-Bench Hard}. 
These are derived from the GSM8K-Aug-NL~\citep{wu25_gsm8k_augnl} and BIG-Bench Hard~\citep{srivastava23_bigbench} corpora. 
The former includes natural language math questions that require multiple steps of arithmetic reasoning. The latter comprises logical reasoning tasks spanning four categories: \texttt{formal\_fallacies}, \texttt{navigate}, \texttt{object\_counting}, and \texttt{web\_of\_lies}. 
We show examples of each task in Table~\ref{tab:task_examples}.
In addition to these single-turn datasets, we construct \emph{Overlapped Spoken GSM8K}, consisting of 161{,}198 conversations (2{,}184 hours).
We generate this dataset by randomly drawing and overlapping two QA pairs from the original \emph{Spoken GSM8K} corpus.

\begin{table}[t]
\centering
\setlength{\belowcaptionskip}{6pt}
\caption{\textbf{Detailed statistics for the constructed training datasets.} Values represent the mean\,$\pm$\,s.d. Both Time-to-First-Response (TTFR) and Time-to-Answer (TTA) are measured from the end of the user's question.}
\label{tab:corpus_budgets}
\footnotesize
\setlength{\tabcolsep}{5pt}
\begin{tabular}{@{}lrr@{}}
\toprule
& Spoken GSM8K & Spoken BIG-Bench Hard \\
\midrule
Source benchmark & GSM8K-Aug-NL & BIG-Bench Hard \\
 & {\scriptsize \citep{wu25_gsm8k_augnl}} & {\scriptsize \citep{srivastava23_bigbench}} \\
\# of Conversations & 365,691 & 54,330 \\
Hours of audio & 2,474 & 667 \\
Length of Reasoning Trace (tok) & 221\,{\scriptsize$\pm$\,200} & 399\,{\scriptsize$\pm$\,236} \\
\midrule
\multicolumn{3}{@{}l}{\emph{$\mathcal{B}=1$ layout}} \\
\quad conversations & 160,398\,{\scriptsize(43.9\%)} & 26,210\,{\scriptsize(48.2\%)} \\
\quad duration / conv. (s) & 32.3\,{\scriptsize$\pm$\,11.1} & 56.0\,{\scriptsize$\pm$\,27.1} \\
\quad TTFR (s) & 1.18\,{\scriptsize$\pm$\,0.17} & 1.16\,{\scriptsize$\pm$\,0.20} \\
\quad TTA (s) & 15.61\,{\scriptsize$\pm$\,8.15} & 29.59\,{\scriptsize$\pm$\,15.27} \\
\addlinespace[2pt]
\multicolumn{3}{@{}l}{\emph{$\mathcal{B}=\infty$ layout}} \\
\quad conversations & 205,293\,{\scriptsize(56.1\%)} & 28,120\,{\scriptsize(51.8\%)} \\
\quad duration / conv. (s) & 18.1\,{\scriptsize$\pm$\,4.8} & 33.2\,{\scriptsize$\pm$\,17.5} \\
\quad TTFR (s) & 1.29\,{\scriptsize$\pm$\,0.18} & 1.27\,{\scriptsize$\pm$\,0.17} \\
\quad TTA (s) & 1.29\,{\scriptsize$\pm$\,0.18} & 1.27\,{\scriptsize$\pm$\,0.17} \\
\bottomrule
\end{tabular}
\end{table}

\begin{table}[p]
\centering
\setlength{\belowcaptionskip}{6pt}
\caption{\textbf{Examples of Spoken Reasoning Tasks} We show one example for each task across \emph{Spoken GSM8K} and \emph{Spoken BIG-Bench Hard}. The reasoning traces come from Gemma4's internal reasoning chain.}
\label{tab:task_examples}
\small
\setlength{\tabcolsep}{4pt}
\begin{tabular}{@{}l p{0.86\textwidth}@{}}
\toprule
\multicolumn{2}{@{}l}{\textbf{Spoken GSM8K}} \\
\addlinespace[1pt]
\emph{Question} & What is ten less than half of 60? \\
\addlinespace[1pt]
\emph{Reasoning} & Here's the step-by-step calculation: \newline 1. Find half of 60: $60 / 2 = 30$. \newline 2. Find ten less than that number (30): $30 - 10 = 20$. \\
\addlinespace[1pt]
\emph{Answer} & 20 \\
\midrule
\multicolumn{2}{@{}l}{\textbf{Spoken BIG-Bench Hard: \texttt{formal\_fallacies}}} \\
\addlinespace[1pt]
\emph{Question} & Here comes a perfectly valid argument: First, being an ancestor of Erik is sufficient for being a schoolmate of Lawrence. Second, Pedro is an ancestor of Erik. From this follows: Pedro is a schoolmate of Lawrence. Is the argument, given the explicitly stated premises, deductively valid or invalid? Options: valid, invalid. Answer the question. \\
\addlinespace[1pt]
\emph{Reasoning} & The user wants me to evaluate the deductive validity of a given argument. \newline Argument Structure: \newline Premise 1: Being an ancestor of Erik $\rightarrow$ Being a schoolmate of Lawrence (A $\rightarrow$ B) \newline Premise 2: Pedro is an ancestor of Erik (A) \newline Conclusion: Pedro is a schoolmate of Lawrence (B) \newline This structure is: \newline 1. $A \rightarrow B$ \newline 2. $A$ \newline $\therefore B$ \newline This is the form of Modus Ponens, which is a deductively valid argument form. \newline Therefore, the argument is valid. \\
\addlinespace[1pt]
\emph{Answer} & valid \\
\midrule
\multicolumn{2}{@{}l}{\textbf{Spoken BIG-Bench Hard: \texttt{navigate}}} \\
\addlinespace[1pt]
\emph{Question} & If you follow these instructions, do you return to the starting point? Take eight steps. Take two steps. Options: Yes, No. Answer the question. \\
\addlinespace[1pt]
\emph{Reasoning} & The user is asking a reasoning question based on a sequence of instructions: \newline 1. Start at a point (let's call it 0). \newline 2. Take eight steps. (New position: 0 + 8 = 8) \newline 3. Take two steps. (New position: 8 + 2 = 10) \newline 4. The question asks: ``do you return to the starting point?'' (Is 10 equal to 0?) \newline Since 10 is not equal to 0, the answer is No. \\
\addlinespace[1pt]
\emph{Answer} & No \\
\midrule
\multicolumn{2}{@{}l}{\textbf{Spoken BIG-Bench Hard: \texttt{object\_counting}}} \\
\addlinespace[1pt]
\emph{Question} & I have a goat, and three cows. How many animals do I have? Answer the question. \\
\addlinespace[1pt]
\emph{Reasoning} & The user is asking a simple addition problem: \newline Goat (1) + Cows (3) = Total animals. \newline 1 + 3 = 4. \\
\addlinespace[1pt]
\emph{Answer} & 4 \\
\midrule
\multicolumn{2}{@{}l}{\textbf{Spoken BIG-Bench Hard: \texttt{web\_of\_lies}}} \\
\addlinespace[1pt]
\emph{Question} & Question: Phoebe lies. Yoland says Phoebe tells the truth. Jamey says Yoland lies. Does Jamey tell the truth? Answer the question. \\
\addlinespace[1pt]
\emph{Reasoning} & Let $P$ be the statement ``Phoebe tells the truth''. \newline Phoebe lies, so $\neg P$ is true. \newline Yoland says ``Phoebe tells the truth''. So, Yoland claims $P$. \newline Since we know $\neg P$ is true, Yoland is lying. \newline Jamey says ``Yoland lies''. \newline We found that Yoland is lying. \newline Therefore, Jamey's statement is true. \newline The question asks: Does Jamey tell the truth? \newline Yes, Jamey tells the truth. \\
\addlinespace[1pt]
\emph{Answer} & Yes \\
\bottomrule
\end{tabular}
\end{table}

\textbf{Construction Details for Single-Turn Datasets.} 
We first employ DeepSeek-V4-Pro~\citep{deepseekv4} to perform text normalization on the questions and answers from the source datasets. Rather than using the originally provided reasoning traces, we prompt Gemma-4 E4B-IT~\citep{gemmateam26_gemma4} with the same questions and record its generated outputs. 
We adopt this approach because training on Gemma's native reasoning traces preserves the model's reasoning capabilities significantly better than relying on out-of-domain traces from the source text. 
We validate these generated traces to ensure logical correctness. 
Next, we synthesize the user questions and system answers into audio using Qwen3 TTS~\citep{Qwen3-TTS}. 
To guarantee the accuracy of this synthesis, we run Parakeet-TDT-0.6b-v3~\citep{Sekoyan25_Canary1Bv2P} on the generated speech. 
Finally, when concatenating the audio segments, we insert a natural conversational gap between the end of the user's question and the moment the system takes the floor.

\textbf{Construction Details for the Overlapped Corpus.} 
We build \emph{Overlapped Spoken GSM8K} using examples from the previously constructed \emph{Spoken GSM8K}. 
Each training instance consists of two randomly selected QA pairs, denoted as ($Q_1, R_1, A_1$) and ($Q_2, R_2, A_2$) corresponding to the questions, reasoning traces, and final answers, alongside any optional spoken fillers ($F_1, F_2$). 
To rigorously test the model's capacity for simultaneous speaking, listening, and reasoning, we restrict the first QA pair to the $\mathcal{B}=1$ configuration. 
This ensures the system actively outputs the filler $F_1$ while processing the reasoning trace $R_1$. 
We then intentionally overlap the audio of $Q_2$ so that the user's second question begins during this exact filler segment. This design forces the model to comprehend incoming audio ($Q_2$) while concurrently speaking ($F_1$) and thinking ($R_1$). 
Because of this specific temporal requirement, $Q_1$ is strictly sampled from $\mathcal{B}=1$ examples, whereas $Q_2$ has no budget constraints.

\section{Training and decoding hyperparameters}
\label{sec:hyper_params}
\textbf{Implementation Details.}
We reuse Gemma's text embeddings and prediction head for \ut{},\st{} and \tta{}.
Text tokenizer is extended with \texttt{<word\_pad>}, \texttt{<silence\_pad>} and \texttt{<empty\_think>}.
Audio embeddings and the prediction head are initialized from scratch.

\textbf{Objective.}
We calculate the total loss as a weighted sum over per-stream cross-entropies, $\mathcal{L} = \sum_{s} \lambda_s \mathcal{L}_s$ with $\lambda_{\ut} = 0.3$, $\lambda_{\st} = 0.5$, $\lambda_{\sa} = 1.0$ and $\lambda_{\tta} = 0.5$.
For \ut{} and \st{}, we downplay the weights of word- and silence-padding tokens to $0.1$ due to their unbalanced occurrence.
For \tta{}, \texttt{<empty\_think>} is down-weighted to $0.1$.

\textbf{Optimization.} We optimize the model using AdamW with $\beta = (0.9, 0.95)$, a weight decay of $0.1$, and a peak learning rate of $2\times10^{-4}$. The learning rate schedule includes $2\,000$ warm-up steps followed by cosine annealing down to $10\%$ of the peak value. 
To preserve stability, the text frontend of the backbone utilizes a reduced learning rate of $2\times10^{-5}$. 
Gradients are clipped at a maximum norm of $1.0$. 
We conduct all training in \texttt{bfloat16} precision using Fully Sharded Data Parallel (FSDP) across two nodes, each equipped with eight NVIDIA H200 GPUs.

\textbf{Decoding Hyperparameters.} 
Each free-running stream is sampled using its own distinct temperature: $0.75$ for \sa{}, $0.7$ for \ut{}, $0.6$ for \st{}, and $0.7$ for \tta{}. All streams employ nucleus sampling with a top-$p$ threshold of $0.95$.

\section{Multi-clock KV caching at decode}
\label{sec:multi_kv}
In this section, we benchmark the decoding speed of our multi-KV cache implementation for \rfull{} and \rgated{}. We evaluate only \rfull{} because both variants share an identical attention shape. 
We report the GPU wall-clock time across various KV cache sizes. As shown in Table~\ref{tab:kv_speed}, the naive wide cache implementation begins to deviate from Single-Clock performance as the KV cache length grows. More importantly, this wide cache approach forces the head dimension to multiply by $C$ as the number of clocks increases, severely degrading efficiency. Conversely, our optimized multi-KV implementation maintains speeds strictly on par with the Single-Clock baseline. These results demonstrate that the multi-KV approach successfully eliminates the decoding speed penalty, allowing us to deploy \rfull{} and \rgated{} without sacrificing hardware efficiency.

\begin{table}[t]
\centering
\setlength{\belowcaptionskip}{6pt}
\caption{\textbf{Decode speed comparison.}
Wall GPU time per decode step in microseconds~($\mu$s) benchmarked with batch size 1 on NVIDIA H200 using FlashInfer~\citep{ye25_flashinfer}.}
\label{tab:kv_speed}
\footnotesize
\setlength{\tabcolsep}{4pt}
\begin{tabular}{@{}lcccccc@{}}
\toprule
 & \multicolumn{6}{c}{KV Cache Length (toks)} \\
\cmidrule(lr){2-7}
 & 512 & 1{,}024 & 2{,}048 & 4{,}096 & 8{,}192 & 16{,}384 \\
\midrule
\multicolumn{7}{@{}l}{\emph{Head Dim $=$ 256}} \\
\quad \singleclock{} & 60.6\,{\scriptsize$\pm$\,0.2} & 61.0\,{\scriptsize$\pm$\,0.2} & 61.8\,{\scriptsize$\pm$\,0.2} & 63.2\,{\scriptsize$\pm$\,0.4} & 69.0\,{\scriptsize$\pm$\,0.3} & 70.7\,{\scriptsize$\pm$\,0.3} \\
\quad \rfull{}~(multi-KV) & 64.4\,{\scriptsize$\pm$\,0.2} & 62.2\,{\scriptsize$\pm$\,0.2} & 62.9\,{\scriptsize$\pm$\,0.3} & 64.5\,{\scriptsize$\pm$\,0.3} & 70.0\,{\scriptsize$\pm$\,0.3} & 72.4\,{\scriptsize$\pm$\,0.8} \\
\quad \rfull{}~(wide cache) & 62.9\,{\scriptsize$\pm$\,0.2} & 63.5\,{\scriptsize$\pm$\,0.3} & 64.0\,{\scriptsize$\pm$\,0.3} & 65.7\,{\scriptsize$\pm$\,0.3} & 72.7\,{\scriptsize$\pm$\,0.9} & 83.0\,{\scriptsize$\pm$\,0.5} \\
\addlinespace[2pt]
\multicolumn{7}{@{}l}{\emph{Head Dim $=$ 512}} \\
\quad \singleclock{} & 63.8\,{\scriptsize$\pm$\,0.2} & 64.3\,{\scriptsize$\pm$\,0.4} & 65.5\,{\scriptsize$\pm$\,0.5} & 68.2\,{\scriptsize$\pm$\,0.6} & 73.2\,{\scriptsize$\pm$\,0.6} & 90.3\,{\scriptsize$\pm$\,0.2} \\
\quad \rfull{}~(multi-KV) & 65.4\,{\scriptsize$\pm$\,0.3} & 63.0\,{\scriptsize$\pm$\,0.3} & 64.1\,{\scriptsize$\pm$\,0.4} & 66.9\,{\scriptsize$\pm$\,0.5} & 72.3\,{\scriptsize$\pm$\,0.3} & 91.7\,{\scriptsize$\pm$\,1.1} \\
\bottomrule
\end{tabular}

\end{table}

\section{Additional budget settings}
\label{sec:extra_budgets}
\begin{table}[t]
\centering
\setlength{\belowcaptionskip}{6pt}
\caption{\textbf{Extreme Thinking Budget results on three benchmarks.}
We report $\mathcal{B}\sim \mathbf{U}\left[1,16\right]$ alongside with other fixed thinking budgets. We include $\mathcal{B}=8$, as this can be viewed it has the same expected thinking budget as $\mathcal{B}\sim \mathbf{U}\left[1,16\right]$.}
\label{tab:extra_budgets}
\scriptsize
\setlength{\tabcolsep}{2.5pt}
\begin{tabular}{@{}lcccccccccccc@{}}
\toprule
 & \multicolumn{4}{c}{Spoken GSM8K} & \multicolumn{4}{c}{BIG-Bench Audio} & \multicolumn{4}{c}{Overlapped} \\
\cmidrule(lr){2-5}\cmidrule(lr){6-9}\cmidrule(lr){10-13}
 & $\mathcal{B}=1$ & $\mathcal{B}=8$ & $\mathcal{B}=\infty$ & R[1,16] & $\mathcal{B}=1$ & $\mathcal{B}=8$ & $\mathcal{B}=\infty$ & R[1,16] & $\mathcal{B}=1$ & $\mathcal{B}=8$ & $\mathcal{B}=\infty$ & R[1,16] \\
\midrule
\multicolumn{13}{@{}l}{\textit{Accuracy}} \\
Single-Clock & 46.7 & 12.6 & 50.0 & 15.5 & 56.6 & 42.3 & 71.4 & 47.0 & 24.0 & 3.0 & 28.1 & 2.8 \\
RUBATO-Split & 44.4 & 42.2 & 45.3 & 42.7 & 77.1 & 76.5 & 80.6 & 76.2 & 21.1 & 3.9 & 17.7 & 4.6 \\
RUBATO-Gated & 47.9 & 49.0 & 48.1 & 48.8 & 78.3 & 79.5 & 79.2 & 78.9 & 20.4 & 10.7 & 17.4 & 12.3 \\
RUBATO-Full & 49.3 & 49.8 & 51.1 & 52.7 & 87.1 & 88.8 & 86.1 & 88.9 & 24.0 & 21.4 & 25.3 & 18.8 \\
\midrule
\multicolumn{13}{@{}l}{\textit{Time to answer}} \\
Single-Clock & 17.4 & 3.5 & 1.5 & 3.4 & 25.7 & 8.0 & 5.7 & 7.5 & 17.0 & 3.5 & 1.0 & 3.7 \\
RUBATO-Split & 17.4 & 4.4 & 1.5 & 4.2 & 23.7 & 7.9 & 5.4 & 7.7 & 17.4 & 3.9 & 1.0 & 3.8 \\
RUBATO-Gated & 17.9 & 5.9 & 1.5 & 5.8 & 22.1 & 10.8 & 7.2 & 10.1 & 17.6 & 3.7 & 1.0 & 4.0 \\
RUBATO-Full & 17.4 & 4.8 & 1.5 & 4.8 & 22.9 & 6.2 & 1.7 & 6.1 & 18.1 & 5.7 & 1.0 & 5.1 \\
\bottomrule
\end{tabular}

\vspace{2pt}
\end{table}

Beyond the budgets evaluated in the main text (Section~\ref{sec:evaluation}), we also test a highly erratic random budget configuration, $R[1, 16]$. 
As shown in Table~\ref{tab:extra_budgets}, performance under this extreme budget perturbation perfectly aligns with the trends observed in our primary results. 
\rubato{} remains significantly more robust than the \singleclock{} baseline, and \rfull{} consistently emerges as the most resilient variant across all conditions. 
This further demonstrates the structural effectiveness of the \rubato{} framework.

\section{Decoding speed of common GPUs}
\label{sec:gpu_budget}
\begin{table}[t]
\centering
\setlength{\belowcaptionskip}{6pt}
\caption{\textbf{Estimated thinking budgets $\mathcal{B}$ for common NVIDIA GPUs.} We estimate the decoding speed and resulting thinking budget for our Gemma-4 E4B-IT backbone and a smaller Llama 3.2 1B model assuming a batch size of 1. The $\mathcal{B}$ columns reflect theoretical maximums based on pure vendor arithmetic. The $\mathcal{B}_{75}$ columns represent a conservative budget assuming 75\% Model Bandwidth Utilization, matching the 65\% to 85\% range observed in real-world deployments. Unmarked bandwidths originate from primary NVIDIA product pages. We denote secondary sources consistent with vendor claims using $^{\dagger}$.}
\label{tab:gpu_budget}
\footnotesize
\setlength{\tabcolsep}{5pt}
\begin{tabular}{@{}lr rrr rrr@{}}
\toprule
& & \multicolumn{3}{c}{Gemma-4 E4B-IT} & \multicolumn{3}{c}{Llama 3.2 1B} \\
\cmidrule(lr){3-5}\cmidrule(lr){6-8}
GPU & BW (GB/s) & tok/s & $\mathcal{B}$ & $\mathcal{B}_{75}$ & tok/s & $\mathcal{B}$ & $\mathcal{B}_{75}$ \\
\midrule
\multicolumn{8}{@{}l}{\emph{Data center}} \\
B200$^{\dagger}$ & 8,000 & 861 & 64 & 47 & 3236 & 254 & 190 \\
H200 SXM$^{\dagger}$ & 4,800 & 517 & 37 & 27 & 1942 & 151 & 112 \\
H100 NVL & 3,900 & 420 & 29 & 21 & 1578 & 122 & 90 \\
H100 SXM & 3,350 & 361 & 24 & 17 & 1355 & 104 & 77 \\
A100 80GB SXM & 2,039 & 220 & 13 & 9 & 825 & 61 & 45 \\
H100 PCIe$^{\dagger}$ & 2,000 & 215 & 13 & 8 & 809 & 60 & 44 \\
A100 80GB PCIe & 1,935 & 208 & 12 & 8 & 783 & 58 & 42 \\
L40S & 864 & 93 & 3 & 1 & 350 & 23 & 16 \\
\addlinespace[2pt]
\multicolumn{8}{@{}l}{\emph{Consumer}} \\
RTX 4090$^{\dagger}$ & 1,008 & 109 & 4 & 2 & 408 & 28 & 20 \\
RTX 3090 Ti$^{\dagger}$ & 1,008 & 109 & 4 & 2 & 408 & 28 & 20 \\
RTX 3090$^{\dagger}$ & 936 & 101 & 4 & 2 & 379 & 26 & 18 \\
\bottomrule
\end{tabular}

\end{table}

In Section~\ref{sec:bg_full-duplex}, we define the thinking budget $\mathcal{B}$ as the number of thinking tokens the model can decode within the duration of a single codec frame. In practice, this value depends on several system parameters: the model backbone size, the codec frame rate, the inference framework, and the underlying hardware. In the main text, to focus purely on model design and empirical evaluation, we abstract these factors into a single budget parameter $\mathcal{B}$ and evaluate our models directly across assigned values. In this section, we provide estimated values of $\mathcal{B}$ across common GPUs to give readers a concrete sense of what realistic deployment budgets look like.

Because hardware setups vary widely in real-world deployments, we report estimated values of $\mathcal{B}$ derived from hardware datasheets (Table~\ref{tab:gpu_budget}). For each GPU, we calculate a theoretical upper-bound estimate alongside a practical estimate assuming 75\% memory bandwidth utilization. We also include numbers for smaller models (e.g., LLaMA 3.2 1B) to contextualize how $\mathcal{B}$ varies with model scale. As shown in the table, sweeping $\mathcal{B}$ from 1 to 32 in our experimental setup covers the decoding throughput of most current GPU configurations.

\textbf{How the estimate is derived.} At a batch size of one without speculative decoding, emitting a token requires reading every parameter touched during the forward pass, making the decoding process memory-bandwidth bound rather than compute-bound. For instance, Gemma-4 E4B-IT accesses approximately $4.64\,\text{B}$ parameters per token. Within an $80\,\text{ms}$ frame, the maximum token capacity is given by $(\text{BW} / \text{GB}_{\text{read}}) \times 0.08$. Of these, four tokens are consumed every frame as a fixed overhead regardless of the budget—one forward pass each for \ua{}, \sa{}, \ut{}, and \st{}. The effective thinking budget $\mathcal{B}$ is the remaining token capacity after deducting this fixed cost.

\section{The mixing ratio of the two training layouts}
\label{sec:ablation_filler}
\begin{figure}[t]
\centering
\includegraphics[width=\textwidth]{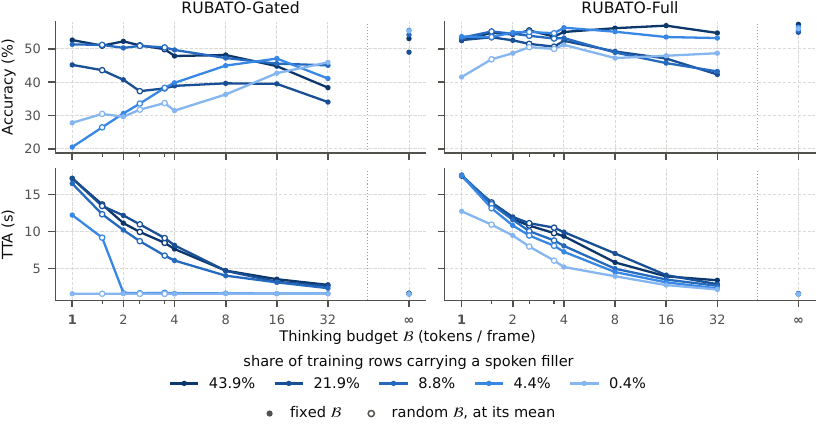}
\caption{\textbf{Mixing proportions of the two training layouts.} Each curve represents a training corpus with a varying proportion of $\mathcal{B}=1$ data (43.9\% to 0.4\%). Following Fig.~\ref{fig:single_overlap}, bold ticks mark the trained budgets, while filled and hollow markers indicate fixed and random per-frame budgets. 
}
\label{fig:filler_ablation}
\end{figure}

As described in Sec.~\ref{sec:two_extremes}, we build every conversation using either a $\mathcal{B}=1$ or $\mathcal{B}=\infty$ budget in roughly equal proportions. This ratio is an adjustable parameter with asymmetric computational costs.
Preparing $\mathcal{B}=\infty$ rows is nearly cost-free, while $\mathcal{B}=1$ rows require synthesizing audio fillers. Consequently, the most economical corpus minimizes $\mathcal{B}=1$ data. Fig.~\ref{fig:filler_ablation} evaluates how aggressively we can reduce this data without degrading interpolation performance.

\textbf{\rfull{} remains robust across altered training proportions compared to \rgated{}.} Across all five dataset configurations, \rfull{} accuracy at $\mathcal{B}=1$ spans a tight 41.6\% to 53.7\%, while \rgated{} fluctuates massively between 20.5\% and 52.7\%. This exact trend persists across all interpolated rates. The \rgated{} model trained at the 0.4\% ratio highlights this instability perfectly. Its time-to-answer flatlines at 1.6\,s across all budgets, proving it abandons the reasoning trace prematurely to force an answer.

\textbf{Accuracy does not decay monotonically as the $\mathcal{B}=1$ data ratio shrinks.} For \rfull{}, training with a 4.4\% share yields performance comparable to a 43.9\% share (53.3\% to 56.4\% versus 52.5\% to 57.0\%). Paradoxically, the intermediate 21.9\% and 8.8\% ratios perform worse and rapidly decay above $\mathcal{B}=4$. Overall, these results indicate that a roughly equal mixture is the safest empirical choice.

\section{Qualitative examples}
\label{sec:qualitative_examples}
Fig.~\ref{fig:qualitative} shows two conversations decoded at $\mathcal{B}=8$, a rate none of these models was trained at, by all four architectures from the same sweep.

\begin{figure}[!ht]
\centering
\begin{minipage}{\textwidth}
\fontsize{7.5}{9}\selectfont
\qlegend{}\par\vspace{3pt}
\noindent\textbf{Question.} Two cups of flour are needed to make a dozen cookies. Carla is making thirty-six cookies today and thirty cookies tomorrow. How many cups of flour will Carla need to bake the cookies today and tomorrow?\quad\textbf{Reference answer:} 11\par\vspace{2pt}
\qcase{qbadbg}{\singleclock{}}{\xmark}{thinks 0.6--2.3\,s, speaks 2.7--39.0\,s}%
  {Here's the plan \qbad{plan}: 1. Determine the flour needed for one dozen cookies (12. Determine the total number of \qbad{of} cookies Carla is making \,\dots\, cups of flour. (implied by the problem context, 2 cups. Carla \qbad{needs44} cups.}%
  {\qnone{} Got it. The answer is 144 cups.}
\qcase{qbadbg}{\rsplit{}}{\xmark}{thinks 0.7--2.2\,s, speaks 0.9--6.0\,s}%
  {Here's the plan: 1. Determine the flour needed for one dozen cookies. (2 cups) 2. Calculate the total number of cookies Carla is making \,\dots\, 66 cookies * 2 cups/cookie = 132 cups. The final answer must be a plain number.}%
  {\qthink{Mmm, right.}\qclose{} Diggy thing. Got it. The answer is one hundred thirty-two cups.}
\qcase{qbadbg}{\rgated{}}{\xmark}{thinks 0.9--2.5\,s, speaks 1.1--25.8\,s}%
  {Here's the plan: 1. Determine the flour needed per dozen cookies. 2. Calculate the total number of cookies needed today and tomorrow. 3. \,\dots\, flour = 66 / 2 Total flour = 33 cups. The final answer must be a plain number.}%
  {\qthink{Bear with me. So two}\qclose{} cups of four per dozen. Thirty-six today, thirty tomorrow. Four needed. Got it. The answer is thirty-three cups.}
\qcase{qokbg}{\rfull{}}{\cmark}{thinks 0.9--4.1\,s, speaks 1.1--8.2\,s}%
  {Here's the plan: 1. Determine the total number of cookies Carla is making. 2. Determine the ratio of flour to cookies (2 cups per dozen). \,\dots\, is 11 cups. The final output must be a number ending with the specific line.}%
  {\qthink{All right, two cups of flour for a dozen.}\qclose{} You want the total? Got it. The answer is eleven cups.}
\vspace{3pt}
\noindent\textbf{Question.} A hospital sees five hundred people a day. Each patient is seen for an average of twenty-four minutes. The doctors charge one hundred fifty dollars an hour to the hospital and the hospital charges the patients two hundred dollars an hour. How much profit does the hospital make from these visits?\quad\textbf{Reference answer:} 10000\par\vspace{2pt}
\qcase{qbadbg}{\singleclock{}}{\xmark}{thinks 0.9--5.4\,s, speaks 5.6--39.7\,s, still going at the 40\,s limit}%
  {Here's the plan step-by-step calculation: 1. **Calculate the total time day:** * 505 hours/day * 240 minutes/hour = 12 minutes/day. * \,\dots\, 150 - \$100 * Profit = \$5 8. **Final Answer \qbad{Formatting5} (as a plain number).**}%
  {\qnone{} Got it. The answer is five hundred dollars.}
\qcase{qbadbg}{\rsplit{}}{\xmark}{thinks 0.8--4.2\,s, speaks 1.1--39.7\,s, still going at the 40\,s limit}%
  {Here's the plan to solve the problem: 1. Calculate the total cost of hospital fees. 2. Calculate the total revenue from hospital visits. 3. \,\dots\, number, no units, no words, no currency symbol, no commas. Profit = -300000.}%
  {\qthink{Right. Five per hour, five hundred thousand}\qclose{} dollars. Got it. The answer is negative three hundred thousand dollars.}
\qcase{qbadbg}{\rgated{}}{\xmark}{thinks 2.1--5.0\,s, speaks 2.3--30.5\,s}%
  {Here's the plan to solve the problem: 1. Calculate the total time spent seeing all patients. 2. Calculate the total revenue from all \,\dots\, Total Revenue - Total Costs Profit = \$200 - \$200 Profit = \$0\$ The profit is 0.}%
  {\qthink{Okay, let me think about this for just a second.}\qclose{} I want to get it right. Got it. The answer is zero At the answer is zero.}
\qcase{qokbg}{\rfull{}}{\cmark}{thinks 0.9--4.4\,s, speaks 1.2--10.9\,s}%
  {Here's the plan to solve the problem: 1. Calculate the total time spent at the hospital. 2. Calculate the total revenue generated by the \,\dots\, = \$10,000\$ The question asks for the profit amount. The profit is 10,000.}%
  {\qthink{Okay, so the hospital charges \$150 an hour.}\qclose{} They were looking at 500 patients. You want a noble profit? Got it. The answer is \$10,000.}
\end{minipage}
\caption{\textbf{Conversations decoded at an unseen budget ($\mathcal{B}=8$).} \emph{Thinks} represents the \tta{} stream and \emph{Says} provides the ASR transcript of the \sa{} stream (any garbled words are ASR errors). \qthink{Teal italic} marks speech generated concurrently with thinking, while the horizontal rule indicates where the reasoning trace concludes. Panels are tinted by the final verdict, with \qbad{highlighting} marking token-level trace defects. The baselines exhibit distinct failure modes. \singleclock{} corrupts its thinking stream mid-arithmetic and remains silent until the trace closes. Meanwhile, \rsplit{} and \rgated{} maintain clean reasoning traces but still calculate the arithmetic incorrectly.}
\label{fig:qualitative}
\end{figure}

\section{Limitations}
\label{sec:limitations}
\textbf{Synthetic audio in experimental setups.} Following the precedent of prior spoken reasoning literature, our experiments utilize synthetic audio. This choice is necessitated by the absence of publicly available full-duplex human conversation datasets featuring time-aligned reasoning traces. While synthetic audio differs from real-world acoustics, the core contribution of \rubato{} centers strictly on the temporal alignment mechanism between independent streams. Because this decoupled clock architecture operates entirely independently of the underlying acoustic features, the structural benefits demonstrated in our evaluation should generalize to real audio recordings.

\textbf{Scope of reasoning tasks and stream count.} The \rubato{} framework is theoretically designed to orchestrate an arbitrary number of interacting modalities and thinking processes. In this work, we scope our empirical validation specifically to a spoken reasoning setup comprising four real-time streams alongside a single chain-of-thought thinking stream. While this configuration shows the efficacy of our temporal decoupling, expanding this validation to include tool use tokens or massively multi-stream environments remains an area for future work pending the availability of suitable datasets.

\textbf{Architectural and codec uniformity.} All empirical results rely on a Gemma-4 E4B-IT backbone paired with NanoCodec operating at 12.5\,Hz using a two-clock configuration. We intentionally fix these architectural variables across all evaluations to maintain a strictly controlled environment. This uniformity successfully isolates the impact of the temporal framework, providing an unambiguous comparison between the statically synchronized baselines and our proposed \rubato{} variants.

\end{document}